\documentclass[aps,pra,twocolumn,amsmath,amssymb,showpacs]{revtex4}

\usepackage{amssymb}
\usepackage{amsmath}
\usepackage{amscd}
\usepackage{graphicx}
\usepackage{natbib}

\newcommand{\be}{\begin{equation}}
\newcommand{\ee}{\end{equation}}

\newcommand{\ab}{\hspace{-1.5mm}}

\newcommand{\g}{\tilde{g}}

\newcommand{\R}{\tilde{R}}

\renewcommand{\vec}[1]{{\mathbf #1}}

\begin{document}

%%%%%%%%%%%%%%%%%%%%%%%%%%%%%%%%%%%%%%%%%%%%%%%%%%%%%%%%%%%%%%%%%%%%%%

\title{
Local atom number fluctuations in quantum gases at finite temperature}

\author{M. Klawunn$^1$, A. Recati$^1$, L. P. Pitaevskii$^{1,2}$, and S. Stringari$^1$}
\affiliation{
\mbox{$^1$INO-CNR BEC Center and Dipartimento di Fisica, Universit\`a di Trento, 38123 Povo, Italy}\\
\mbox{$^2$Kapitza Institute for Physical Problems, Russian Academy of Science, Kosygin street 2, 119334 Moscow, Russia}}

\date{\today}
 
\begin{abstract}
%We study the number fluctuations of quantum gases in a finite volume at finite temperature. 
%We show that fluctuations measured in a small volume differ significantly from the fluctuations obtained from the 
%fluctuation-dissipation theorem. We consider in detail free Fermi gases and weakly interacting Bose gases with and without dipolar interaction.
We investigate the number fluctuations in small cells of quantum gases pointing out  
important deviations from the thermodynamic limit fixed by the isothermal compressibility. Both quantum and thermal fluctuations in weakly 
as well as  highly compressible fluids are considered. For the 2D superfluid Bose gas we find a significant quenching of fluctuations with respect to the thermodynamic limit, in agreement with recent experimental findings. An enhancement of the thermal fluctuations is instead predicted for the 2D dipolar superfluid Bose gas, which becomes dramatic when the size of the sample cell is of the order of the wavelength of the rotonic excitation induced by the interaction.

\end{abstract}


\maketitle

\section{Introduction}

The  experimental possibility of detecting few and even single atoms confined in magnetic or optical traps with high precision is opening new perspectives in the study of  correlations \cite{WestbrookCorr,Blochcorr} and  fluctuations \cite{Westbrook1D,Esslinger,Ketterle2010,Bouchoule,Chin2010,Cornell2D} in atomic and molecular gases at the mesoscopic and microscopic scale.

In the present work we investigate the fluctuations of the particle number in small sample cells of quantum gases, exploring the deviations from the predictions of thermodynamic theory  which relates the fluctuations to the isothermal compressibility. 

We consider two important examples of continuous systems. The first one is the ideal Fermi gas, a benchmark  of statistical mechanics, which has been recently studied experimentally in atomic gases  to point out the anti-bunching effect produced by the Pauli exclusion principle \cite{Esslinger,Ketterle2010}. The second system  is the dilute quasi-two dimensional (2D) bosonic gas in the superfluid regime. Two dimensional gases are particularly suited to measure the number fluctuations since one avoids the column density integration. Measurements of this kind have been recently carried out showing important quenching of the fluctuations in the degenerate regime\cite{Chin2010}, whose origin is clearly explained by our analysis. On the other hand quasi-2D Bose gases with dipolar interaction present an enhancement of number fluctuations. This is due to the roton-like spectrum \cite{Goraroton,roton-like} which has important consequences on the behaviour of the static structure factor, thus on the number fluctuations, especially at finite temperature.

When the size of the sample cell is large enough one can use the thermodynamic theory of fluctuations (see, e.g., \cite{LL5}). 
The expression for particle number fluctuations at temperature $T$ reads
\be
\langle \delta N^2 \rangle= N k_B T n\kappa_T,
\label{eq:FDT}
\ee
which relates the fluctuations to the isothermal compressibility
$\kappa_T=1/n^2\,\partial n /\partial \mu\equiv 1/nmc^2$, with $n$ the average density, $\mu$ the chemical potential and $c$ the speed of sound \cite{c}.
When $T$ is large the compressibility of the gas approaches the classical value $n \kappa_T=1/k_B T$
and one recovers the shot noise regime. An extension of the thermodynamic relation for the density fluctuations
of non-uniform systems has been recently proposed as a tool to measure temperatures \cite{HoTerm}. 
If the radius of the cell is not large enough,
important deviations with respect to Eq. (\ref{eq:FDT}) start occurring.
In our investigation we will mainly focus on the low temperature regime 
where quantum effects become particularly important. 

The paper is organized as follows. In the next section we recall
how to calculate the particle number fluctuations from the correlation function and
the static structure factor, respectively. In section III we discuss the temperature dependence
of fluctuations in a finite volume of an ideal Fermi gas in three dimensions.
In the same section we also study the dependence of quantum fluctuations ($T=0$)
on the shape of the probe volume. Then in section IV we analyse the 
density fluctuations for the case of an interacting quasi-2D Bose gas at temperatures
far below the critical temperature. We condsider
short-range interactions as well as long-range dipole-dipole interactions
and discuss differences in the static structure factor and in the finite volume fluctuations.
In the same section we provide an analytical expression -- whose derivation
for completeness is given in the Appendix -- for the zero-temperature fluctuations in 2D and for a contact potential,

\section{Fluctuations in a finite volume}

The particle number fluctuations $\langle \delta N^2 \rangle$ in a finite cell of volume $V$ are calculated by  double integration of the density-density 
correlation function $ n\nu(|\vec{r}_1-\vec{r}_2|)=\langle n(\vec{r}_1)n(\vec{r}_2)\rangle-{n}^2$ over $V$ and can be written as\cite{SLS98,Astra2007}
\begin{align}\label{fluc_r}
\langle \delta N^2 \rangle=
n\ab\int \nu(\vec{r}) h(\vec{r}) d\vec{r}=
n \ab\int \!S(\vec{k}) H(\vec{k}) \frac{d\vec{k}}{(2\pi)^D},
\end{align}
where $D$ is the dimensionality of the system,
$h(\vec{r})=\int_V\int_V\delta(\vec{r}_1-\vec{r}_2-\vec{r}) d\vec{r}_1 d\vec{r}_2$
is a geometrical factor which depends on the shape of the probe cell, $H(\vec{k})=\int_V\int_V e^{i\vec{k}(\vec{r}_1-\vec{r}_2)} d\vec{r}_1 d\vec{r}_2$ is its Fourier transform and $ S(\vec{k})=\int  \nu(\vec{r}) e^{-i\vec{k}\vec{r}}   d\vec{r}$ is the static structure factor. Geometrically $h({\bf r})$ corresponds to the volume of the overlapping region between a cell of volume $V$ and the same cell shifted by the vector ${\bf r}$.
For large volumes (thermodynamic limit) $H({\bf k})\rightarrow V\delta({\bf k})$ and one finds $\langle\delta N^2 \rangle=NS(k=0)$ yielding, at finite temperature, the thermodynamic expression Eq.(\ref{eq:FDT}). Deviations from the thermodynamic limit are directly related to the finite-$k$ behaviour of the static structure factor. In order to discuss these deviations it is useful to distinguish between weakly compressible and highly compressible fluids.
Weakly compressible fluids are characterized by the condition $ mc^2 \ge E_{deg}$ where $E_{deg} \sim \hbar^2n^{2/D}/m$ is the degeneracy energy and include systems like the ideal Fermi gas, the unitary Fermi gas and superfluid Helium.
For such systems the low temperature condition $k_B T \ll mc^2$ is easily reachable, yielding antibunching with respect to the shot noise limit, i.e., $\delta N^2 < N$.
Highly compressible fluids are instead characterized by the condition  $mc^2 \ll E_{deg}$.  For such systems, which include dilute superfluid Bose gases, the low temperature condition $ k_B T \ll mc^2$ is hardly reachable experimentally and the fluctuations naturally exhibit bunching effects with respect to the shot-noise limit. 
In the following we will consider two important examples: the ideal Fermi gas and the 2D Bose gas.

\section{Ideal Fermi Gas}

As a paradigmatic example of weakly compressible fluids we consider the ideal Fermi gas, whose finite temperature density-density correlation function in three dimensions reads
\begin{align}\label{nu_Fermi}
\ab\nu(\vec{r})\!=\!\delta(\vec{r})\!-\ab
\left[  \frac{k_B T}{n \epsilon_F}\frac{k_F^2}{2\pi^2 r} \int_0^\infty \!\!\ab du
\frac{u \sin\left(u \sqrt{\frac{k_B T}{\epsilon_F}} k_F r \right)}{e^{-\mu(T) /k_B T}\,e^{u^2}+1} \right]^2\ab\ab,
\end{align}
where $k_F=(6\pi^2 n)^{1/3}$ and $\epsilon_F=\hbar^2 k_F^2/2m$ are the Fermi momentum and the Fermi energy, respectively.
The chemical potential of the ideal Fermi gas $\mu(T)$ is obtained self-consistently from
\begin{align}\label{mu_T_Fermi}
\left(\frac{k_B T}{\epsilon_F} \right)^{3/2} \int _0^\infty \!\!\ab du
\frac{3 u^2}{e^{-\mu(T) /k_B T}\,e^{u^2}+1} 
=1
\end{align}

Initially, we consider the particle number fluctuations in a spherical cell of radius $R$, for which the geometrical factor is
\begin{align}
h(r)=\frac{4\pi R^3}{3}-\pi r R^2+\frac{\pi r^3}{12},
\end{align}
with $r\in[0,2 R]$.
In Fig. \ref{fig:fluc_T} we show $\langle \delta N^2 \rangle$ as a
function of temperature for  $N=4/3 \pi R^3 n=100$ and $2000$ calculated from Eq. (\ref{fluc_r}) (solid lines)
and from Eq. (\ref{eq:FDT}) (dashed line). As expected the two results are the closer the larger the volume (and hence $N$)  and, for finite volumes,
they approach each other at large temperature when the deBroglie wavelength becomes smaller than the sample size.
In particular the fluctuations in a finite volume are always larger then the ones predicted by the thermodynamic expression Eq. (\ref{eq:FDT}) and remain finite at $T=0$ where they have a purely quantum nature. In this limit,
to the leading order in $N$, they can be written as \cite{Astra2007,Yvan,Recati2010}
$\langle \delta N^2 \rangle \approx 3 (3/(32\pi^4))^{1/3} N^{2/3} \ln(10.45 N^{1/3})$. Such an expression suggests 
that quantum fluctuations have a surface nature being dominated by the dependence $N^{2/3}\propto R^2$.
Let us also note that the approximation
$\langle \delta N^2 \rangle \approx n N k_B \kappa_{T=0} =3 N k_B T/(2\epsilon_F)$ obtained by neglecting the $T$-dependence in the compressibility (see  dashed-dotted line in Fig. \ref{fig:fluc_T}), 
is bad for all $T$: at low temperatures quantum fluctuations are not taken into account
and at large temperature the relation $\kappa_T\approx\kappa_{T=0}$ is not valid.

\subsection{Surface shape dependence of quantum fluctuations}

\begin{figure}[ht]
\begin{center}
\includegraphics[width=0.42\textwidth,angle=0]{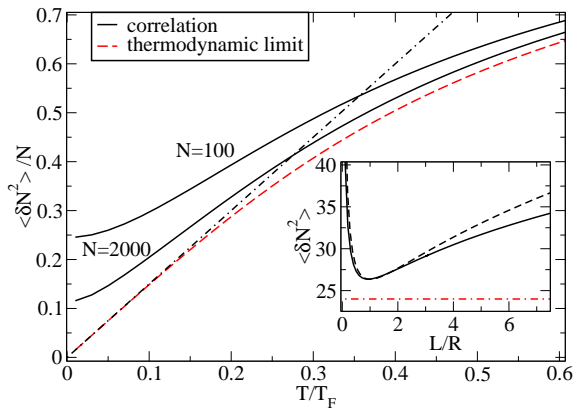}
\caption{Number fluctuation $\langle \delta N^2 \rangle$ for a free Fermi gas as a function of $T/T_F$ calculated 
integrating the density-density correlations (solid line) for $N=100$ and $N=2000$ and using the thermodynamic expression Eq. (\ref{eq:FDT}) (dashed line).
Inset: $\langle \delta N^2 \rangle$ at $T=0$ for a cylinder (solid line) and for a sphere (dashed-dotted line) 
for $N=100$ as a function of the aspect ratio $L/R$. The fluctuations closely follow the cylinder surface dependence (dashed line) for a fixed volume (see text).}
\label{fig:fluc_T}
\end{center}

\end{figure}

In order to get a better insight into the geometrical dependence of the quantum fluctuations \cite{Ketterle_private} we calculate $\langle \delta N^2 \rangle$ at $T=0$ for a cylinder with radius $R$ and length $L$, keeping the volume (and hence $N$) fixed. 
The particle number fluctuations are given by Eq.~(\ref{fluc_r}),
with the geometrical factor
\begin{align}
h(\rho,z)\!=\! (2 L\!-\!z)\left[2 R^2 \arccos\left(\frac{\rho}{2R}\right)\!-\!\frac{\rho}{2}\sqrt{4 R^2-\rho^2}\right]\ab,
\end{align}
where $z\in[0, 2L]$ and $\rho\in[0,2 R]$.
The results are shown in the inset of Fig. \ref{fig:fluc_T}  as a function of the aspect ratio $L/R$ and for $N=100$.
We find that the particle number fluctuations are minimal for $L\simeq R$ where the surface  is minimal and they are even smaller for a sphere 
with the same volume (horizontal dashed-dotted line).
Interestingly, in the inset of Fig. \ref{fig:fluc_T} we also show that the result follows closely the dependence  of the surface  
$S_{Cyl}=2 V^{2/3}/\pi^{1/3}[2(L/R)^{1/3}+(R/L)^{2/3}]$ of the cylinder on the ratio $L/R$, for a fixed volume $V$. Deviations from the latter expression is due to the logarithmic term, which cannot be taken into account by the geometrical analysis.

\vspace{0.5cm}
%%%%%%%%%%%%
% FIGURE fluc_T
%%%%%%%%%%%%%%

\section{2D Bose gas with dipolar interaction}

As important examples of highly compressible fluids we analyse dilute bosonic gases in the superfluid regime. 
Moreover we consider the system to be in a quasi-2D geometry where the motion along $z$ is frozen. 
This is usually realised with a strong harmonic confinement in the $z$ direction. 
We consider a quasi-2D gas, where the three dimensional $s$-wave scattering length $a$ is much smaller than
the harmonic oscillator length $l_z$ in the $z$ direction. In this case it is possible to define an effective 2D short-range coupling constant 
$g=\hbar^2\sqrt{ 8\pi } a/(m l_z)$. We will also include  a long range dipole-dipole interaction, characterised by 
the parameter $g_d=2\sqrt{8\pi} d^2/(3 l_z)$, where $d$ is the dipole moment of the atom (or the molecule) which we take 
oriented perpendicular to the two dimensional plane of the gas.

Let us introduce the healing length $\xi=\hbar/(mc)$ and the chemical
potential $\mu=mc^2=\hbar^2/(m \xi^2) =g(1+\beta) n$, where $\beta=g_d/g=2md^2/(3\hbar^2 a)$ is the ratio between 
of the dipole-dipole and the contact interaction strengths\cite{Feshbach}. Within Bogoliubov theory the spectrum of elementary excitations is \cite{roton-like}
\begin{align}\label{spectrum}
\frac{\epsilon(k)}{mc^2}= \xi k \sqrt{\frac{\xi^2 k^2}{4}+1-
\frac{\beta}{1+\beta}\frac{3}{2 \sqrt{\pi}} F \left(\frac{k l_z}{\sqrt{2}}\right)},
\end{align}
with $F(x)=x \,\mathrm{erfc}\left( x \right)e^{x^2}$.
In the absence of dipolar interaction ($\beta=0$) this spectrum has the well-known Bogoliubov form.
For finite $\beta$ the dipolar term can significantly change the shape of the spectrum.
For negative $s$-wave scattering length $a$ the spectrum
can exhibit a roton-like minimum. Note, that in this case the gas must be stabilised against phonon instability, which demands $|\beta|>1$. 

\subsection{Static structure factor}
At low temperatures $T \ll T_{\lambda}$, where the gas is superfluid, the static structure factor $S(k,T)$ is related to the dispersion relation $\epsilon(k)$
of the elementary excitations via
\begin{equation}\label{Sk}
S(k,T)=\frac{\hbar^2 k^2}{2 m \epsilon(k)} \coth \frac{\epsilon(k)}{2 k_B T}.
\end{equation} 
The latter relation, Eq. (\ref{Sk}), follows from the fact that the imaginary part of the response function in a weakly interacting Bose gas
is $T$-independent and takes the form of a $\delta$-function \cite{BECbook}.

\begin{figure}[ht]
\vspace{0.5cm}
\begin{center}
\includegraphics[width=0.4\textwidth,angle=0]{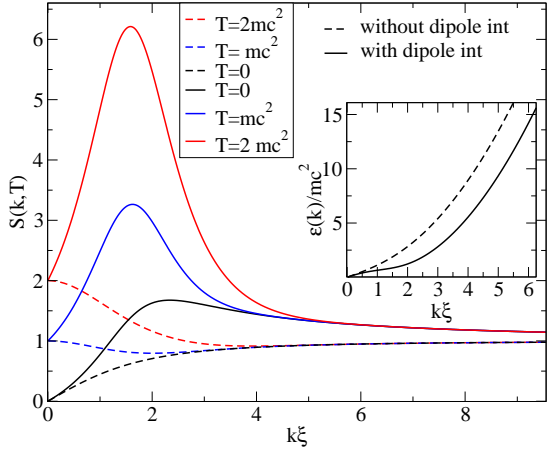}
\caption{Static structure factor $S(k,T)$ for a quasi-2D Bose gas with (continuous lines) and without (dashed lines) dipolar interaction for $k_B T/mc^2=0$, $1$, $2$.
Inset: the dispersion relation Eq. (\ref{spectrum}). For the dipolar case we use values for $^{52}$Cr atoms, namely $a=-1.5a_0$, $l_z=0.09\mu m$ which corresponds to $\beta=-1.083$, $n/(\sqrt{2\pi} l_z)=10^{14}$ cm$^{-3}$ and $\xi=2.5\mu$m.}
\label{fig:Sk}
\end{center}
%\vspace*{-0.2cm}
\end{figure}

The static structure factor depends significantly on the ratio $k_B T/(mc^2)$. In Fig. \ref{fig:Sk} 
we show $S(k,T)$ as a function of $k\xi$ for different temperature with and without dipolar interaction.
The peculiar roton-like shape of $\epsilon(k)$ shows up in a maximum in $S(k)$. This effect is significantly amplified at finite temperature as can be seen in 
Fig. \ref{fig:Sk}. Actually, even if the spectrum does not exhibt a roton minimum (see the inset in the figure), a clear peak is visible in the static structure factor.
Physically, this is due to thermal occupation of the roton-like states at finite $k$. The enhancement can be easily understood from Eq. (\ref{Sk}) since, if $\epsilon(k)\ll k_B T$, the static form factor reduces to
\be
S(k,T)\simeq S(k,0){2 k_B T\over \epsilon(k)}={\hbar^2k^2 k_B T\over m\epsilon(k)^2},
\label{SkhighT}
\ee 
and in particular at low momenta we have  $S(k,T)\simeq (1+A k\xi)k_BT/(mc^2)$ for $g_d\neq 0$, with $A=3 \beta l_z/[\sqrt{8\pi}(1+\beta)]$ and $S(k,T)=(1-(k\xi/2)^2)k_BT/(mc^2)$ for $g_d=0$. It is worth noticing that the large thermal enhancement of the the structure factor near the roton minimimum is 
peculiar of dilute gases. In fact in a weakly compressible system, like superfluid helium the thermal energy $k_B T$ is always smaller than the roton energy. Moreover Bragg spectroscopy, a technique typically used to extract $S(k)$ in quantum gases, 
can only access the $T=0$ static structure factor, being sensitive to the imaginary part of the response function, rather than to the dynamic structure factor \cite{BECbook}.
We will show that  the study of the particle number fluctuations can instead reveal the effect of the sizable temperature enhancement in $S(k)$  produced by the rotonic excitation.

% FIGURE Sk

\subsection{Fluctuations without dipolar interactions}

\begin{figure}[ht]
%\begin{center}
\includegraphics[width=0.42\textwidth,angle=270]{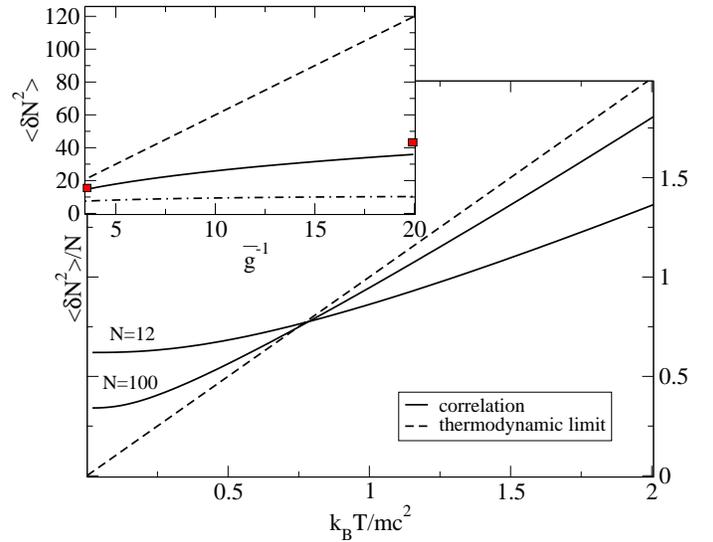}
\caption{$\langle \delta N^2 \rangle/N$ for a quasi-2D Bose gas in the superfluid regime  as a function of $k_B T/(mc^2)$
for $N=12$, $100$ and $g=\hbar^2/4 m$, calculated from Eq. (\ref{fluc_r}) (solid lines) and from Eq. (\ref{eq:FDT}) (dashed lines).
Inset: $\langle \delta N^2 \rangle$ as a function of the inverse of the dimensionless coupling constant $\bar{g}=mg/\hbar^2$ for $N=12$, $T=20nK$ calculated 
from  Eq. (\ref{fluc_r}) (solid lines) and from Eq.(\ref{eq:FDT}) (dashed lines). The (red) small boxes are the experimental value extracted from \cite{Chin2010} at $T\approx 20$nK. The $T=0$ result Eq. (\ref{eq:T02D}) is also reported for comparison (dashed-dotted line).}
\label{fig:fluc_T_Bose}
\end{figure}

First, we study the temperature dependence of the fluctuations
in the non-dipolar case ($g_d=0$).
In order to calculate the particle number fluctuations in a disk of radius $R$
we use Eq.  (\ref{fluc_r}) with the geometrical factor 
\begin{align}\label{H_2D}
H_{2D}(k)={4\pi^2 R^2 J_1^2(R k)\over k^2},
\end{align}
where $ J_1$ denotes the Bessel function.
%First, we study the temperature dependence of the fluctuations
%in the non-dipolar case.
At $T=0$ the fluctuations can be calculated analytically (see appendix A)
\be\label{eq:T02D}
\langle \delta N^2 \rangle_{T=0}=\frac{\hbar^2}{m g}\frac{R}{\xi} \ln\left[ C \frac{R}{\xi} \right],
\ee
with $C \approx 7.4$. The result for $\delta N^2$ as a function of temperature and for $N=12$, $100$ is shown in Fig. \ref{fig:fluc_T_Bose} where we also report the result (\ref{eq:FDT}), which holds in the thermodynamic limit. 
We used the parameters of the experiment reported in Ref. \cite{Chin2010}. In the inset we plot the fluctuations versus the interaction strength and compare with values measured in the superfluid regime of experiment \cite{Chin2010}. The agreement is pretty good. Further we show that the fluctuations are closer to the $T=0$ result Eq. (\ref{eq:T02D}) (dashed-dotted line) than to the thermodynamic limit.
At finite $T$ we calculate the fluctuations using Eq. (\ref{Sk}) with the dispersion relation Eq. (\ref{spectrum}) for $\beta=0$.
% FIGURE fluc_T_Bose
\vspace{0.4cm}

Fig. \ref{fig:fluc_T_Bose} shows that we have a significant deviation from the thermodynamic relation Eq. (\ref{eq:FDT}) at all temperatures. 
We can explain the behaviour of the fluctuation in an easy way within a simple approximation which 
describes how the fluctuations are related to the size of the sample cell.
indeed if we consider a sphere of radius $R$ the relevant momenta contributing to the integral (\ref{fluc_r}) are  of the order $R^{-1}$ and  the fluctuations can be approximated by the expression 
$\delta N^2\sim NS(\alpha/R,T)$, where $\alpha$ is a constant of order unity. 
There are  three natural length scales in the problem: the healing length $\xi$,
the phonon thermal wavelength $l_p=\hbar c/k_B T$ and the de Broglie wavelength $l_T=\hbar/\sqrt{m k_B T}$.

If $R\gg\xi$ the static structure factor reduces to
\be
S\left(k\xi,\frac{k_B T}{mc^2}\right)\simeq{\alpha\xi \over {2 R}}\coth\left(\frac{mc^2}{2k_BT}{\alpha\xi\over R}\right)={\alpha\xi \over{2 R}}\coth\left({\alpha\over 2}{l_p\over R}\right),
\ee
and, as expected, only the phonon part of the spectrum plays a role.
If the radius is also larger than the phonon thermal wavelength ($R\gg l_p$) we get,
as expected \cite{LL5}, $\delta N^2=N k_B T/mc^2$, i.e., we recover the thermodynamic result and, depending on the value of $T/mc^2$, the fluctuations can be sub- or super-poissonian.
If, instead, $R\ll l_p$ we have $\delta N^2=\alpha N \xi/2R\propto N^{1/2}$, i.e., aside from the log-term,
we recover the $T=0$ quantum result.
Notice that the latter regime requires the condition $k_BT\ll mc^2$ and hence is difficult to access in dilute Bose gases.

In the opposite $R\ll\xi$ regime, we expect to probe the particle-like part of the spectrum. Indeed the static structure factor reads
\be
\! S\!\left(k\xi,\frac{k_B T}{mc^2}\right)
\!\simeq  \coth\!\left[\frac{mc^2}{4k_B T}\!\left({\alpha\xi\over R}\!\right)^2\!\right]\!=\!\coth\!\left[\left(\alpha{l_T\over R}\right)^2\right]\ab.
\ee
Again we can distinguish two cases:
for $R\gg l_T$, which is possible only for $k_B T\gg mc^2$, 
we get $\delta N^2=N k_B T/mc^2 (R/\xi)^2$, i.e., the fluctuations
are reduced with respect to the thermodynamic limit,  as clearly shown in Fig. \ref{fig:fluc_T_Bose} for $k_BT/mc^2>1$.
If instead $R\ll l_T$, i.e. if $R$ is the smallest length scale of the problem, we recover the shot-noise result $\delta N^2=N$ which,by the way,  coincides with the $T=0$ ideal Bose gas result.
\vspace{0.5cm}
\begin{figure}[ht]
\includegraphics[width=0.4\textwidth,angle=0]{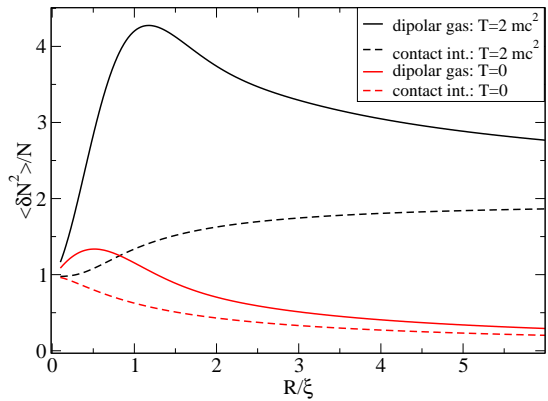}
\caption{$\langle \delta N^2 \rangle/N$ as a function of $R/\xi$ at zero temperature (dashed lines) and $k_B T/(mc^2)=2$ (solid lines)
for a dipolar gas of $^{52}$Cr atoms (black lines) and for a non-dipolar gas $(\beta=0)$ (grey lines)
with the same chemical potential. Parameters as in Fig. \ref{fig:Sk} and in particular $\xi=2.5\mu$m.}
\label{fig:fluc_T_DipolevsR}
%\end{center}
\end{figure}

\subsection{Fluctuations with dipolar interactions}

In order to calculate the particle number fluctuations in the presence of dipolar interactions $(g_d\neq0)$
in a disk of radius $R$ we use again Eq.  (\ref{fluc_r}) with the geometrical factor \ref{H_2D}.
In the previous subsection we have seen that a discussion of the proper length scales
can give a good estimation for the behaviour of the flutuations.  
For a dipolar gas there is another important length scale in the problem: $1/k_{max}$,
which corresponds to the maximum of $S(k)$. 
Notice that even a small deviation of the excitation spectrum
from the usual Bogoliubov form, without exhibiting a minimum (see inset in Fig. \ref{fig:Sk}) 
leads to the maximum in $S(k)$, which is strongly amplified at finite $T$ as shown by Eq. (\ref{SkhighT}).
Hence, for sample cells of size $R\sim 1/k_{max}$ we expect, at finite $T$,
a significant amplification of the particle number fluctuations. In Fig. \ref{fig:fluc_T_DipolevsR}, we report $\delta N^2$ calculated 
for a dipolar gas of $^{52}$Cr atoms under experimentally feasible conditions \cite{Muller}. 
Notice that the amplification of the fluctuations survive also for larger 
values of the cell size and 
%Contrary to the appearance of a roton-like minimum in the spectrum,
%the amplification of fluctuations is robust with respect to uncertanties
%in experimental parameters (e.g. in the negative $s$-wave scattering length
%for Cr$^52$), because it follows directly from the maximum in $S(k)$.
the measurement of the particle number fluctuations is expected to provide a very sensitive tool
to reveal the  temperature amplification
of the static structure factor in the rotonic-like region.

\section{Conclusion}

We have shown that the investigation of the atom number fluctuations at the local scale provides 
a new insight on the microscopic structure of the correlations present in quantum gases where quantum and thermal 
effects combine in a non-trivial way.
In particular, we explain the quenching of atom number fluctuations measured in a quasi-2D superfluid Bose gas \cite{Chin2010}. In a quasi-2D dipolar gas we obtain a crucial thermal enhancement of fluctuations.
%the presence of rotonic excitations.
This opens new perspectives for future experimental investigations.
The present analysis can be naturally extended to investigate also spin-fluctuations in quantum systems \cite{Recati2010,KetterleSpin}.

A more quantitative comparison with experiments, where the cell $V$ has not sharp boundaries, would require the inclusion of the proper smooth weight function in the convolution integral Eq. (\ref{fluc_r}). 

\acknowledgements We acknowledge very useful discussion with C. Chin, T. Esslinger, W. Ketterle, S. M\"uller and G. Shlyapnikov. This work has been supported by ERC through the QGBE grant.

\appendix

\section{Zero temperature fluctuations of the 2D Bose-gas}

In the present appendix we calculate  analytically the particle number fluctuations at 
$T=0$ from Eq. (\ref{fluc_r}) for a disk
with radius $R$. Since the spectrum is given by Eq. (\ref{spectrum}) for $\beta=0$, and the geometrical factor by Eq. (\ref{H_2D}), the integral giving the number fluctuations can be written as 
\begin{equation}\label{fluc_T0}
\langle \delta N^2 \rangle= N  \int_0^\infty \ab d k \,
\frac{J_1^2(R k/\xi)}{\sqrt{1+\frac{k^2}{4}}}.
\end{equation} 
%with $N$ the number of particles.

\vspace{0.5cm}
% FIGURE fluc_T0
\begin{figure}[ht]
\begin{center}
\includegraphics[width=0.42\textwidth,angle=0]{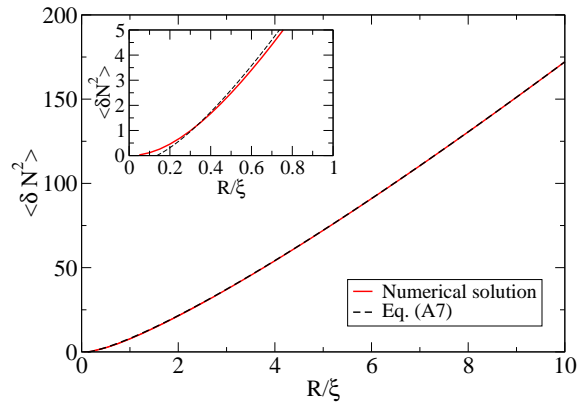}
\caption{Comparison of $\langle \delta N^2 \rangle$ as a function of $R/\xi$ for $\g=1/4$
calculated numerically from Eq. (\ref{fluc_T0}) and Eq. (\ref{fluc_T=0_a}).
Inset: Zoom onto small $R/\xi$.}
\label{fig:fluc_T0}
\end{center}
\end{figure}

In order to get an analytical result we use the property of products of Bessel functions \cite{Eason1954,Watsonbook}
%implied by the addition theorems of Bessel functions (see e.g. \cite{Watsonbook}). 
\begin{equation}
J_{\eta+n}(A) J_n(B)= \frac{1}{\pi} \int_0^\pi \ab d\theta  \,
\frac{J_\eta(X)}{X^\eta}\left(  A-B e^{i \theta}\right)^\eta e^{-in\theta},
\end{equation} 
with $X^2=A^2+B^2-2AB \cos (\theta)$, which in our case reads
\begin{equation}
J^2_1(\R k)=\frac{1}{\pi} \int_0^\pi \ab d\theta \,
J_0\left[ 2 R/\xi k \sin \left( \frac{\theta}{2}\right) \right]  e^{-i\theta}.
\end{equation} 
We introduce this in Eq. (\ref{fluc_T0}) and 
interchange the integrations
\begin{equation}
\langle \delta N^2 \rangle= \frac{N}{\pi} \int_0^\pi \ab d\theta  \,
F(\theta) e^{-i\theta}.
\end{equation} 
Then we are able to calculate the inner integral
\begin{align}
F(\theta) &\equiv 2 \int_0^\infty \ab dy 
\frac{J_0\left[ 4 R/\xi \sin \left( \frac{\theta}{2}\right) y \right]}{\sqrt{1+y^2}}\\
 &=2 \, I_0\left[ 2 R/\xi \sin \left( \frac{\theta}{2}\right) y \right]
         \, K_0\left[ 2 R/\xi \sin \left( \frac{\theta}{2}\right) y \right],\nonumber
\end{align} 
with the modified Bessel functions of zeroth order $I_0$ and $K_0$.
In order to simplify the integral we split
it at a small cutoff $\epsilon:=2\arcsin(\delta)$, with $\delta=(2R/\xi C_1)^{-1}$
small for sufficiently large $R/\xi$. Here $C_1$ is an unknown numerical constant. 
The cutoff $\epsilon, \delta \ll 1$ is chosen such, that one can use
the expansion of $I_0(z) K_0(z) \approx 1/(2z)$ (for large argument $z$)
at $\theta > \epsilon$.
The equation for the fluctuations simplifies to
\begin{align}
\langle &\delta N^2 \rangle= \frac{2N}{\pi} \left\lbrace 
\int_\epsilon^\pi \ab d\theta  \,
\frac{e^{-i\theta}}{4 R/\xi \sin \left( \frac{\theta}{2}\right) } \right. \\
&\left. + \int_0^\epsilon  \ab d\theta  \, e^{-i\theta}
\, I_0\!\left[ 2 R/\xi \sin \left( \frac{\theta}{2}\right) y \right]
         \, K_0\!\left[ 2 R/\xi \sin \left( \frac{\theta}{2}\right) y \right]
\! \right\rbrace \nonumber
\end{align} 
One can show that the second integral vanishes for $\epsilon \ll 1$,
in spite of the logarithmic divergence of $K_0$ at small arguments.
Carrying out the first integral for $\epsilon, \delta \ll 1$
we finally obtain the particle number fluctuations
\begin{equation}\label{fluc_T=0_a}
\langle \delta N^2 \rangle=\frac{R/\xi}{\g} \ln\left[ C R/\xi \right],
\end{equation} 
with $C:= 2C_1e^{-2+\ln 2}$.
By numerically calculating the integral we find $C\approx 7.4$.
This corresponds to $C_1\approx13.7$ and thus $\delta\approx(27 R/\xi)^{-1}$.
Hence  $\epsilon, \delta \ll 1$ is satisfied as long as $R/\xi\gg 0.04$.
Fig. (\ref{fig:fluc_T0}) compares Eq. (\ref{fluc_T=0_a}) with the numerical solution 
of the integral in Eq. (\ref{fluc_T0}).

%\bibliography{bibliothek}

\begin{thebibliography}{99}
\bibitem{c}
Here $c$ is the isothermal speed of sound which at low temperature is indistinguishable from the adiabatic one.
\bibitem{Feshbach}
In this system the value of the scattering, and then $\beta$, can be tuned by means of Feshbach resonance \cite{Cr-Fesh}.
\end{thebibliography}

\end{document}